\documentclass[12pt]{article}
\usepackage{graphicx}

\newsavebox{\sboxpubnumber}
\newsavebox{\sboxpubdate}
\newcommand{\pubdate}[1]{\begin{lrbox}{\sboxpubdate}{#1}\end{lrbox}}

\newcommand{\Title}[1]{\begin{center} {\Large #1 } \end{center}}
\newcommand{\Author}[1]{\begin{center}{ \sc #1} \end{center}}
\newcommand{\Address}[1]{\begin{center}{ \it #1} \end{center}}

\newenvironment{Abstract}{\begin{quotation}  }{\end{quotation}}
\newenvironment{Presented}{\begin{quotation} \begin{center}
             PRESENTED AT\end{center}\bigskip
      \begin{center}\begin{large}}{\end{large}\end{center}
      \end{quotation}}

\newcommand{\CL}{{\cal L}}

\newcommand{\CO}{{\cal O}}

\newcommand{\bear}{\begin{array}}  \newcommand{\eear}{\end{array}}
\newcommand{\bea}{\begin{eqnarray}}  \newcommand{\eea}{\end{eqnarray}}
\newcommand{\beq}{\begin{equation}}  \newcommand{\eeq}{\end{equation}}
\newcommand{\bef}{\begin{figure}}  \newcommand{\eef}{\end{figure}}
\newcommand{\bec}{\begin{center}}  \newcommand{\eec}{\end{center}}
\newcommand{\non}{\nonumber}  
\newcommand{\lmk}{\left(}  \newcommand{\rmk}{\right)}
\newcommand{\lkk}{\left[}  \newcommand{\rkk}{\right]}
\newcommand{\lhk}{\left \{ }  \newcommand{\rhk}{\right \} }

\newcommand{\bib}{\bibitem} 
\newcommand{\la}{\left\langle} \newcommand{\ra}{\right\rangle}

\newcommand{\gtrsim} {~ \raisebox{-1ex}{$\stackrel{\textstyle >}{\sim}$} ~} 
\newcommand{\lesssim} {~ \raisebox{-1ex}{$\stackrel{\textstyle <}{\sim}$} ~}

\def\IBB#1#2#3{{\bf #1}, #2 (20#3)}

\def\AA#1#2#3{Astron. Astrophys. {\bf #1}, #2 (19#3)}

\def\APJ#1#2#3{Astrophys. J. {\bf #1}, #2 (19#3)}
\def\APJL#1#2#3{Astrophys. J. Lett. {\bf #1}, L#2 (19#3)}

\def\MNRAS#1#2#3{Mon. Not. R. Astron. Soc. {\bf #1}, #2 (19#3)}
\def\MPLA#1#2#3{Mod. Phys. Lett. A {\bf #1}, #2 (19#3)}
\def\NAT#1#2#3{Nature (London) {\bf #1}, #2 (19#3)}

\def\NPB#1#2#3{Nucl. Phys. {\bf B#1}, #2 (19#3)}

\def\PLB#1#2#3{Phys. Lett. B {\bf #1}, #2 (19#3)}

\def\PLBold#1#2#3{Phys. Lett. {\bf#1B}, #2 (19#3)}

\def\PRD#1#2#3{Phys. Rev. D {\bf #1}, #2 (19#3)}
\def\PRDD#1#2#3{Phys. Rev. D {\bf #1}, #2 (20#3)}

\def\PRL#1#2#3{Phys. Rev. Lett. {\bf#1}, #2 (19#3)}
\def\PRLL#1#2#3{Phys. Rev. Lett. {\bf#1}, #2 (20#3)}

\def\PTP#1#2#3{Prog. Theor. Phys. {\bf #1}, #2 (19#3)}

\begin{document}

\begin{titlepage}
\pubdate{\today}                    

\vfill
\Title{Double Inflation in Supergravity and Primordial Black Holes
Formation }
\vfill
\Author{Masahide Yamaguchi}
\Address{Research Center for the Early Universe, University of Tokyo,
Tokyo 113-0033, Japan} 
\vfill

\begin{Abstract}
    In this talk we propose a natural double inflation model in
    supergravity. Chaotic inflation first takes place by virtue of the
    Nambu-Goldstone-like shift symmetry. During chaotic inflation, an
    initial value of second inflation (new inflation) is set, which is
    adequately far from the local maximum of the potential due to the
    small linear term in the K\"ahler potential.  Then, primordial
    fluctuations within the present horizon scale may be produced
    during both inflations. Primordial fluctuations responsible for
    anisotropies of the cosmic microwave background radiation and the
    large scale structure are produced during chaotic inflation, while
    fluctuations on smaller scales are produced during new inflation.
    Because of the peculiar nature of new inflation, they can become
    as large as $10^{-1}$-$10^{-2}$, which may lead to the formation
    of primordial black holes.
\end{Abstract}
\vfill
\begin{Presented}
    COSMO-01 \\
    Rovaniemi, Finland, \\
    August 29 -- September 4, 2001
\end{Presented}
\vfill
\end{titlepage}
\def\thefootnote{\fnsymbol{footnote}}
\setcounter{footnote}{0}

\section{Introduction}

\label{sec:int}

Primordial black holes (PBHs) are very interesting. Because massive
compact halo objects (MACHOs) are observed through gravitational
microlensing effects \cite{MACHO}, which are a possible candidate of
dark matter.  Furthermore, PBHs evaporating now may be a source of
antiproton flux observed by the BESS experiment \cite{BESS} or
responsible for short gamma ray bursts (GRBs) \cite{gamma}. PBHs may
be realized in the context of double inflation. Though many double
inflations have been considered, they are often discussed in a simple
toy model with two massive scalar fields. In this talk, we propose a
natural double inflation model in SUGRA, where chaotic inflation takes
place first of all, during which an initial value of new inflation is
dynamically set due to the supergravity effects. It can be adequately
far from the local maximum of the potential due to the small linear
term of the inflaton in the K\"ahler potential. Therefore primordial
density fluctuations responsible for the observable universe can be
attributed to both inflations, that is, chaotic inflation produces
primordial fluctuations on large cosmological scales and new inflation
on smaller scales.\footnote{In Ref. \cite{YY} the initial value of new
inflation is so close to the local maximum of the potential for new
inflation that the universe enters a self-regenerating stage
\cite{eternal,sr}. Therefore primordial fluctuations responsible for
the observable universe are produced only during new inflation.} The
energy scale of new inflation becomes of the same order as the initial
value of new inflation so that produced density fluctuations may
become as large as the order of unity due to the peculiar nature of
new inflation, which straightforwardly may lead to PBHs formation.

\section{Model and dynamics}

\subsection{Model}

In this section we propose a double inflation model in
supergravity\cite{Yama}. We introduce an inflaton chiral superfield
$\Phi(x,\theta)$ and assume that the model, especially, K\"ahler
potential $K(\Phi,\Phi^{\ast},\dots)$ is a function of
$\Phi+\Phi^{\ast}$, which enables the imaginary part of the scalar
component of the superfield $\Phi$ to take a value larger than the
gravitational scale, and leads to chaotic inflation. Such a functional
dependence of $K$ can be attributed to the Nambu-Goldstone-like
symmetry introduced in Ref. \cite{KYY}. We also introduce a spurion
superfield $\Xi$ describing the breaking of the shift symmetry and
extend the shift symmetry as follows,
\bea
  \Phi &\rightarrow& \Phi + i~C M_{G}, \non \\
  \Xi  &\rightarrow& \lmk\frac{\Phi}{\Phi + i~C M_{G}}\rmk^{2} \Xi,
  \label{eq:shift}
\eea
where $C$ is a dimensionless real constant. Below, the reduced Planck
scale $M_{G}$ is set to be unity. Under this shift symmetry, the
combination $\Xi\Phi^{2}$ is invariant. Inserting the vacuum value
into the spurion field, $\la \Xi \ra = \lambda$, softly breaks the
above shift symmetry. Here, the parameter $\lambda$ is fixed with a
value much smaller than unity representing the magnitude of breaking
of the shift symmetry (\ref{eq:shift}).

We further assume that in addition to the shift symmetry, the
superpotential is invariant under the U$(1)_{R}$ symmetry because it
prohibits a constant term in the superpotential. The above K\"ahler
potential is invariant only if the $R$-charge of $\Phi$ is zero. Then,
we are compelled to introduce another supermultiplet $X(x,\theta)$
with its $R$-charge equal to two, which allows the linear term $X$ in
the superpotential. As shown later, for successful inflation, the
absolute magnitude of the coefficient of the linear term $X$ must be
at most of the order of $|\lambda|$, which is much smaller than unity.
Therefore in order to suppress the linear term of $X$ in the
superpotential, we introduce the $Z_{2}$ symmetry and a spurion field
$\Pi$ with odd charge under the $Z_{2}$ symmetry and zero $R$-charge.
The vacuum value $\la \Pi \ra = v$ softly breaks the $Z_{2}$ symmetry
and suppress the linear term of $X$. Charges for superfields are shown
in table I.

After inserting vacuum values of spurion fields $\Xi$ and $\Pi$, and
neglecting higher order terms, the superpotential is given by
\bea
  W &\simeq& vX - \lambda X\Phi^{2} \\
    &=& vX(1 - g\Phi^{2})
\eea
with $g \equiv \lambda / v$. Here we set both constants $v$ and
$\lambda$ ($g$) to be real for simplicity.

The K\"ahler potential neglecting a constant term and higher order
terms is given by
\beq
  K = v_{2}(\Phi + \Phi^{\ast}) 
     + \frac12 (\Phi + \Phi^{\ast})^{2} 
     + XX^{\ast}.
  \label{eq:kahler}
\eeq
Here $v_{2} \sim v$ is a real constant representing the breaking
effect of the $Z_{2}$ symmetry. Here and hereafter, we use the same
characters for scalar with those for corresponding supermultiplets.

\subsection{Dynamics}

Decomposing the scalar field $\Phi$ into real and imaginary
components
\beq
  \Phi = \frac{1}{\sqrt{2}} (\varphi + i \chi),
\eeq
the Lagrangian density $L(\varphi,\chi,X)$ is given by
\beq
  L(\varphi,\chi,X) = 
              \frac{1}{2}\partial_{\mu}\varphi\partial^{\mu}\varphi 
              + \frac{1}{2}\partial_{\mu}\chi\partial^{\mu}\chi 
              + \partial_{\mu}X\partial^{\mu}X^{*}
              -V(\varphi,\chi,X)
\eeq
with the potential $V(\varphi,\chi,X)$ given by
\bea \hspace{-1.0cm}
  V(\varphi,\chi,X)
    &=& v^{2} e^{-\frac{v_{2}^{2}}{2}}
           \exp \lhk \lmk \varphi + \frac{v_{2}}{\sqrt{2}} \rmk^{2} 
                     + |X|^{2}
                \rhk \non \\ 
    && \hspace{0.0cm} \times
         \lhk~\lkk 
              1 - g (\varphi^{2} - \chi^{2}) 
             + \frac14~g^{2} (\varphi^{2} + \chi^{2})^{2}
              \rkk
             (1-|X|^{2}+|X|^{4}) 
         \right. \non \\ 
    && \hspace{0.5cm}
             +~|X|^{2} 
              \lkk~
                2g^{2}(\varphi^{2}+\chi^{2})
              \right. \non \\
    && \hspace{2.0cm}
                - (v_{2}+\sqrt{2}\varphi) \lhk
                  \sqrt{2}~g\varphi
                    \lkk~2 - g (\varphi^{2} - \chi^{2}) 
                    ~\rkk - 2\sqrt{2}~g^{2}\varphi\chi^{2}
                    \rhk \non \\
    && \hspace{2.0cm} \left. \left.    
                 +(v_{2}+\sqrt{2}\varphi)^{2} 
                    \lhk
                      1 - g (\varphi^{2} - \chi^{2}) 
                    + \frac14~g^{2} 
                    ~(\varphi^{2} + \chi^{2})^{2}
                    \rhk
               ~\rkk
                      ~\rhk.
\eea

Because of the exponential factor, $\varphi$ and $X$ rapidly goes down to
$\CO(1)$. On the other hand, $\chi$ can take a value much larger than
unity without costing exponentially large potential energy. Then the
scalar potential is approximated as
\beq
  V \simeq \lambda^{2}
                   \lmk \frac{\chi^{4}}{4}      
                        + 2 \chi^{2} |X|^{2}
                   \rmk
  \label{eq:twopot}
\eeq
with $\lambda = g v$. Thus the term proportional to $\chi^{4}$ becomes
dominant and chaotic inflation can take place. Then, using the
slow-roll approximation, the $e$-fold number $\widetilde{N}$ during
chaotic inflation is given by
\beq
  \widetilde{N} \simeq \frac{\chi_{\widetilde{N}}^{2}}{8}.
  \label{eq:chaoe}
\eeq
The effective mass squared of $\varphi$, $m_{\varphi}^{2}$, during
chaotic inflation becomes
\beq
  m_{\varphi}^{2} \simeq \frac{\lambda^{2}}{2} \chi^{4}
                  \simeq 6H^{2} \gg \frac94 H^{2},
                  ~~~~~~H^{2} \simeq \frac{\lambda^{2}}{12}\chi^{4},
\eeq
where $H$ is the hubble parameter at that time. Therefore $\varphi$
oscillates rapidly around the minimum $\varphi_{min}$ so that its
amplitude damps in proportion to $a^{-3/2}$ with $a$ being the scale
factor. Here, the potential minimum for $\varphi$, $\varphi_{min}$,
during chaotic inflation is given by
\beq
  \varphi_{min} \simeq - v_{2}/\sqrt{2}.
\eeq
Thus the initial value of the inflaton $\varphi$ of second inflation
(new inflation) is set dynamically during chaotic inflation.

On the other hand, the mass squared of $X$, $m_{X}^{2}$, is dominated
by
\beq
  m_{X}^{2} \simeq 2 \lambda^{2} \chi^{2} \simeq \frac{24}{\chi^2}H^2,   
  \label{mhratio}
\eeq
which is much smaller than the hubble parameter squared until
$\chi^{2} \sim 24$ so that $X$ also slow-rolls. However we can easily
show that $X$ is much smaller than unity throughout the chaotic
inflation regime.

As $\chi$ becomes smaller and of order of unity, new inflation starts.
The potential with $X \simeq 0$ is approximated as
\bea
  V(\varphi,\chi,X \simeq 0)
    &\simeq& v^{2} e^{-\frac{v_{2}^{2}}{2}}
           \exp \lmk \varphi + \frac{v_{2}}{\sqrt{2}} 
                \rmk^{2} 
                \non \\ 
    && \hspace{-2.0cm} \times
         \lkk 
             \lmk 
               1 - \frac{g}{2} \varphi^{2} 
             \rmk^{2}
            + \chi^{2}
             \lmk
               g + \frac{g^{2}}{2} \varphi^{2}
                   + \frac{g^{2}}{2} \chi^{2}
             \rmk
         \rkk.  
\eea
The global minima are given by $\varphi^{2} = 2/g$ and $\chi = 0$. The
mass squared for $\varphi$, $m_{\varphi}^{2}$, reads
\beq
  m_{\varphi}^{2} 
      \simeq - (g - 1) + (g + \frac12 g^{2}) \chi^{2}.
\eeq
Thus new inflation begins when $\chi \simeq \chi_{crit}$ given by
\beq
  \chi_{crit} =  \frac{2}{g} 
                        \sqrt{\frac{g - 1}{g + 2}}.
\eeq

Once new inflation begins, $\chi$ rapidly goes to zero because the
effective mass squared becomes $m_{\chi}^{2} \simeq 6 g H^{2} \ge 6
H^{2}$. Then, for $\chi \simeq 0$ and $X \ll 1$, the potential is
given by
\beq
  V(\varphi,\chi \simeq 0,X \ll 1) \sim
    v^{2} \lhk 1 - (g - 1) \widetilde\varphi^{2} 
                 + 2 (g - 1)^{2} \widetilde\varphi^{2}|X|^{2}
                 + \cdots
          \rhk,
\eeq
where $\widetilde\varphi = \varphi - \varphi_{max}$ and $\varphi_{max}
\equiv v_{2}/[\sqrt{2}(g-1)]$. Thus if $g \ge 1$ ($\lambda \ge v$),
new inflation takes place and $\varphi$ rolls down slowly toward the
vacuum expectation value $\eta = \sqrt{2/g}$.

Before new inflation starts, $\varphi$ stays at $\varphi_{min}$, which
is different from $\varphi_{max}$. Then, the initial value of
$\widetilde\varphi$, $\widetilde\varphi_{i}$ for new inflation is given by
\beq
  \widetilde\varphi_{i} = - \frac{v_{2}}{\sqrt{2}}
                         \frac{g}{g-1}.
\eeq
On the other hand, the amplitude of quantum fluctuations of $\varphi$
is estimated as $\delta\varphi_{q} \sim v / (2\pi\sqrt{3})$. Using the
fact that $g \ge 1$, we find that quantum fluctuations do not dominate
the dynamics unless $v_{2} \ll v$.

The total $e$-folding number $N_{new}$ during new inflation is given
by
\beq
  N_{new} \simeq \frac{1}{2(g-1)}
                \ln \left| \frac{\sqrt{2}}{v_{2}}
                           \frac{g-1}{g} \right|.
\eeq
Then the total $e$-folding number $N$ is given by $N = \widetilde{N} +
N_{new}$. We set $N_{\rm COBE} = 60$ for simplicity, when the physical
wavenumber of the mode ($k_{\rm COBE}$) corresponding to the Cosmic
Background Explorer~(COBE) scale exits the horizon, that is, $k_{\rm
COBE}/(a_{(N=60)} H_{(N=60)}) = 1$.

In case $N_{new} \gtrsim 60$, primordial density fluctuations
responsible for the observable universe are produced only during new
inflation. Otherwise, chaotic inflation produces primordial
fluctuations on large cosmological scales and new inflation on smaller
scales. In this paper we consider only the latter case.

After new inflation, $\varphi$ oscillates around the global minimum
$\eta$ and the universe is dominated by a coherent scalar-field
oscillation of $\sigma \equiv \varphi -\eta$. Expanding the
exponential factor $e^{v_{2}\varphi + \varphi^2} $ in $e^K$,
\beq
  e^{v_{2}\varphi + \varphi^2} = e^{\eta^2}(1+2\eta\sigma+\cdots ),
\eeq
we find that $\sigma$ has gravitationally suppressed linear
interactions with all scalar and spinor fields including minimal
supersymmetric standard model (MSSM) particles. For example, let us
consider the Yukawa superpotential $W = y_{i}D_{i}HS_{i}$ in MSSM,
where $D_{i}$ and $S_{i}$ are doublet (singlet) superfields, $H$ is a
Higgs superfield, and $y_{i}$ is a Yukawa coupling constant. Then the
interaction Lagrangian is given by
\beq
  \CL_{\rm int} \sim 
     y_{i}^{2} \eta \sigma D_{i}^{2} S_{i}^{2} + \cdots,
\eeq
which leads to the decay width $\Gamma$ given by 
\beq
  \Gamma \sim \sum_{i} y_{i}^4 \eta^2 m_{\sigma}^3.
  \label{eq:gamma}
\eeq
Here $m_{\sigma} \simeq 2\sqrt{g_{R}}e^{\sqrt{2/g_{R}}}v$ is the mass
of $\sigma$. Thus the reheating temperature $T_{R}$ is given by
\beq
  T_{R} \sim 0.1 \bar{y} \eta m_{\varphi}^{3/2},
\eeq
where $\bar{y}=\sqrt{\sum_{i} y_{i}^4}$. Taking
$\bar{y} \sim 1$, the reheating temperature $T_{R}$ is given by
\beq
  T_{R} \sim v^{3/2} \lesssim \lambda^{3/2}. 
\eeq
As shown later, the upper bound of $\lambda$ is given by $\lambda <
1.2 \times 10^{-6}$. Hence the reheating temperature $T_{R}$ is
constrained as
\beq
  T_{R} \lesssim 10^{-9} \sim 10^{9}~{\rm GeV},
\eeq
which is low enough to avoid the overproduction of gravitinos in a
wide range of the gravitino mass \cite{Ellis}.

\section{Density fluctuations and PBHs formation}

\subsection{Density fluctuations}

In this section we investigate primordial density fluctuations
produced by this double inflation model. First of all we consider
density fluctuations produced during chaotic inflation. As shown in
the previous section, there are two effectively massless fields,
$\chi$ and $X$, during chaotic inflation. However it is easily shown
that only $\chi$ contributes to growing adiabatic fluctuations
\cite{PS}. Then, with the fact that $X \ll 1$, the amplitude of
curvature perturbation $\Phi_{A}$ on the comoving horizon scale at
$\chi=\chi_{\widetilde{N}}$ is given by the standard one-field formula
and reads
\beq
  \Phi_{A}(\widetilde{N})
            \simeq  \frac{f}{2\sqrt{3}\pi}
                  \frac{\lambda\chi_{\widetilde{N}}^{3}}{8}, 
  \label{eq:gpotentialc}
\eeq
where $f=3/5~(2/3)$ in the matter (radiation) domination. If $N_{new}
\lesssim 60$, the comoving scale corresponding to the COBE scale exits
the horizon during chaotic inflation. Defining $\widetilde{N}_{\rm
COBE}$ as the $e$-folding number during chaotic inflation,
corresponding to the COBE scale, the COBE normalization requires
$\Phi_{A}(\widetilde{N}_{\rm COBE}) \simeq 3\times 10^{-5}$
\cite{COBE}. Then the scale $\lambda$ is given by
\beq
    \lambda \simeq 4.2 \times 10^{-3} 
        \chi_{\widetilde{N}_{\rm COBE}}^{-3}. 
    \label{eq:COBEnor}
\eeq
The spectral index $n_{s}$ is given by
\beq
  n_{s} \simeq 1 - \frac{3}{\widetilde{N}_{\rm COBE}}.
\eeq
Since the COBE data shows $n_s = 1.0 \pm 0.2$ \cite{COBE},
$\widetilde{N}_{\rm COBE} \ge 15$, which leads to $\lambda < 1.2
\times 10^{-6}$.

Next let us discuss density fluctuations produced during new
inflation. In this case also, both $\widetilde\varphi$ and $X$ are
effectively massless fields. However, as with the case of chaotic
inflation, it is easily shown that only $\widetilde\varphi$
contributes to growing adiabatic fluctuations. Then, with the fact
that $X \ll 1$, the amplitude of curvature perturbation $\Phi_{A}$ on
the comoving horizon scale at
$\widetilde\varphi=\widetilde\varphi_{N}$ is given by
\beq
  \Phi_{A}(N) \simeq \frac{f}{2\sqrt{3}\pi} 
       \frac{v}{ 2 (g - 1) \widetilde\varphi_{N}}.
  \label{eq:gpotentialn}
\eeq
The spectral index $n_s$ of the density fluctuations is given by
\beq
    n_s \simeq 1 - 4 (g - 1).
\eeq

You should also notice that $\widetilde\varphi \sim v_{2}$ at the
beginning of new inflation. Then, since $v \sim v_{2}$, the amplitude
of curvature perturbation $\Phi_{A}$ can become as large as the order
unity, which may lead to PBHs formation.

\subsection{Primordial black holes formation}

PBHs have been paid renewed attention to because they may explain the
existence of massive compact halo objects (MACHOs) \cite{MACHO} and
become a part of cold dark matter. Furthermore, PBHs are responsible
for antiproton fluxes observed by the BESS experiments \cite{BESS} or
short gamma ray bursts \cite{gamma}.

Carr and Hawking first discussed PBHs formation and showed that in the
radiation dominated universe, a black hole is formed soon after the
perturbed region reenters the horizon if the amplitude of density
fluctuations $\delta$ lies in the range $1/3 \le \delta \le 1$
\cite{CH}. Then, the mass of produced PBHs $M_{BH}$ is roughly given
by the horizon mass,
\beq
  M_{BH} \simeq \frac{4 \sqrt{3} \pi}{\sqrt{\rho}}
         \simeq 0.066 M_{\odot} \lmk \frac{T}{\rm GeV} \rmk^{-2},
  \label{eq:BHmass}
\eeq
where $\rho$ and $T$ are the total energy density and the temperature
of the universe at formation. The horizon scale at formation is
related to the present cosmological scale $L$ by
\bea
  L &\simeq& \frac{a(T_{0})}{a(T)} H^{-1}(T), \non \\
    &\simeq& 6.4 \times 10^{-8} {\rm Mpc} 
       \lmk \frac{T}{\rm GeV} \rmk^{-1},\
  \label{eq:BHscale}
\eea
with $T_{0} \simeq 2.7 $ K the present temperature of the universe.
The corresponding comoving wave number $k = 2\pi / L$ is given by
\beq
  k \simeq 1.0 \times 10^{8} {\rm ~Mpc}^{-1} 
                       \lmk \frac{T}{\rm GeV} \rmk.\
  \label{eq:BHwave}
\eeq

Assuming Gaussian fluctuations, the mass fraction of produced PBHs
($\beta \equiv \rho_{BH}/\rho$) is given by
\bea
  \delta(M) &=& \int^{1}_{1/3} \frac{1}{\sqrt{2\pi}\sigma(M)}
               \exp \lmk - \frac{\delta^{2}}{2 \sigma^{2}(M)} \rmk
                d\delta, \non \\
            &\simeq& \sigma(M) 
               \exp \lmk - \frac{1}{18 \sigma^{2}(M)} \rmk,
  \label{eq:BHfraction}
\eea
where $\sigma(M)$ is the root mean square of mass variance evaluated
at horizon crossing.

Using the mass fraction $\beta$ at formation, the ratio of the present
energy density $\rho_{BH}(M)$ of PBHs with the mass $M$ and the
entropy density is given by
\beq
  \frac{\rho_{BH}(M)}{s} \simeq \frac34 \beta(M) T,
\eeq
which yields the normalized energy density,
\bea
  \Omega_{BH} h^{-2} &\simeq& 2.1 \times 10^{8} \beta 
                           \lmk \frac{T}{\rm GeV} \rmk \non \\   
                     &\simeq& 5.4 \times 10^{7} \beta 
                           \lmk \frac{M}{M_{\odot}} \rmk^{-1/2} \non \\   
                     &\simeq& 2.1 \beta 
                           \lmk \frac{k}{\rm Mpc^{-1}} \rmk.
\eea

Then, MACHO PBHs with mass $\sim 0.1M_{\odot}$ are produced at the
temperature given by
\beq
  T \simeq 0.81 {\rm ~GeV},
\eeq
which corresponds to
\bea
  L &\simeq& 7.9 \times 10^{-8} {\rm ~Mpc}, \non \\
  k &\simeq& 8.1 \times 10^{7} {\rm ~Mpc^{-1}}.
  \label{eq:MACHOscale}
\eea
As easily seen from Eq. (\ref{eq:gpotentialn}), the fluctuations with
the largest amplitude are produced at the onset of new inflation,
which we identify with the formation time of PBHs. The spectrum is so
steep that the formation of the PBHs with smaller masses is suppressed
strongly. The present energy density $\Omega_{BH} h^{-2} \sim 0.25$ is
explained if the mass fraction is given by
\beq
  \beta \simeq 1.5 \times 10^{-9},
\eeq
which implies the mass variance $\sigma \simeq 0.056$ under the
Gaussian approximation, corresponding to
\beq
  \Phi_{A} \sim 0.04.
  \label{eq:MACHOP}
\eeq
Taking into account Eqs. (\ref{eq:MACHOscale}), (\ref{eq:MACHOP}), and
the COBE normalization (\ref{eq:COBEnor}), MACHO PBHs are produced in
this model if we take the parameters given by
\bea
  \lambda &\sim& 1.3 \times 10^{-6}, \non \\
  v &\sim& 1.1 \times 10^{-6}, \non \\
  v_{2} &\sim& 0.93 \times 10^{-6},
\eea 
with $g = \lambda / v \simeq 1.2$.

Note that all parameters are of the same order. Also, the temperatures
at formation are lower than the reheating temperature so our
assumption that PBHs are formed in the radiation dominated universe is
justified.

\section{Discussion and conclusions}

\label{sec:con}

In this talk, we have proposed a natural double inflation model in
SUGRA. By virtue of the shift symmetry, chaotic inflation can take
place, during which the initial value of new inflation is set.  The
initial value of new inflation is adequately far from the local
maximum of the potential so that primordial fluctuations within the
present horizon scale are attributed to both inflations. That is,
fluctuations responsible for the anisotropy of the CMB and the large
scale structure are produced during chaotic inflation, while
fluctuations on smaller scale are produced during new inflation. Due
to the peculiar nature of new inflation, fluctuations on smaller scale
are as large as of the order of unity, which may lead to PBHs
formation. As an example, we consider MACHO PBHs. We find that if we
take reasonable values of parameters, such PBHs are produced in our
double inflation model.

\begin{table}[t]
  \begin{center}
    \begin{tabular}{| c | c | c | c | c |}
                   & $\Phi$ & $X$ & $\Xi$ & $\Pi$ \\
        \hline
        $Q_R$      & 0      & 2   & 0     & 0 \\ 
        \hline 
        $Z_{2}$    & $-$    & $-$ & $-$   & $-$   
    \end{tabular}
    \caption{The charges of various supermultiplets of U$(1)_{R}
    \times Z_{2}$.} 
    \label{tab:charges}
  \end{center}
\end{table}

\end{document}